return

# The Delineation of an Interdisciplinary Specialty in terms of a Journal Set: The Case of Communication Studies

Loet Leydesdorff

Amsterdam School of Communications Research (ASCoR), University of Amsterdam,

Kloveniersburgwal 48, 1012 CX Amsterdam, The Netherlands; loet@leydesdorff.net;

http://www.leydesdorff.net

&

Carole Probst

Faculty of Communication Sciences, Università della Svizzera italiana,

Switzerland; carole.probst@lu.unisi.ch

**Abstract**

A journal set in an interdisciplinary or newly developing area can be determined by including the journals classified under the most relevant ISI Subject Categories into a journal-journal citation matrix. Despite the fuzzy character of borders, factor analysis of the citation patterns enables us to delineate the specific set by discarding the noise. This methodology is illustrated using communication studies as a hybrid development between political science and social psychology. The development can be visualized using animations which support the claim that a specific journal set in communication studies is increasingly developing, notably in the "being cited" patterns. The resulting set of 28 journals in communication studies is smaller and more focused than the 45 journals classified by the ISI Subject Categories as "Communication." The proposed method is tested for its robustness by extending the relevant environments to sets including many more journals.

## Introduction

Almost by definition, interdisciplinary and new developments take place at the interfaces between two or more disciplines and are therefore difficult to capture using *ex ante* classificatory schemes. One can expect citation traffic between disciplines to be less dense than within disciplinary cores. This diminishes the visibility in terms of numbers of citations and thereof derived indicators for statistical reasons. Laudel & Origgi (2006) suggested that interdisciplinary research systematically receives lower grades than disciplinary research efforts in research assessment exercises. Indexing tends to begin with already established delineations, to be cautious about adding new categories, and therefore new developments are incorporated only with a delay.

Morillo *et al*. (2001 and 2003) used co-classifications of the ISI Subject Categories as indicators of the interdisciplinarity of journals (Bordons *et al*., 2004). Van Leeuwen & Tijssen (2000) explored the use of citation traffic among these categories to map the dynamics of interdisciplinary developments. In a recent communication, Leydesdorff & Schank (2008) noted that interdisciplinary developments may initially be diffuse, but then tend to stabilize at the interface between existing specialties. A dynamic perspective can add to the analysis because an emerging density can be distinguished more easily from the perspective of hindsight. From an *ex ante* perspective, the newly emerging density is first interwoven in a co-evolution of previously existing specialties and because of fluctuations between years it may remain less clear whether and when a new identity is to be acknowledged (Leydesdorff, 2002).

## The case of communication studies

In this study, we apply the above reasoning to the field of communication studies. Rogers (1999) argued that communication studies have remained deeply divided between two subdisciplines: mass communication and interpersonal communication. This divide could be retrieved empirically in terms of (a) the lack of cross-citations among the major journals of these two subdisciplines (Rice *et al*., 1988; So, 1988), (b) the structure of the professional associations (Barnett & Danowski, 1992; Doerfel & Barnett, 1999), and (c) the awarding of doctoral degrees in programs specializing in either of the two subfields (Rogers, 1994). Rogers (1999, at p. 618) identified the origins of this "intellectual canyon" as largely historical, but accidental factors have reinforced this split running through the field of communication studies.

Important contributions to the field of communication studies have historically been made by scholars from a wide variety of disciplines such as political science, sociology, psychology, and even mathematics. Historians of communication studies have noted a temptation to rely on ideas from other fields (Beniger, 1993; Putnam, 2001; Schramm, 1983; Streeter, 1995). The field's boundaries, however, have consequently remained unclear. Scholars in communication studies tend to import ideas from other fields more than they export new theories and methods to these other fields (Berger, 1991; Boure, 2006; Reeves & Borgman, 1983; Rice *et al.,* 1988; So, 1988).

Wilbur Schramm, for example, a founding father of the field who wrote a history of its development (Schramm, 1983; cf. Delia 1987; Rogers, 1994), identified "the political scientist, Lasswell; the mathematician-turned-sociologist, Lazarsfeld; the social psychologist and student of group processes, Lewin; and the experimental-turned-social psychologist, Hovland" as the founding fathers of the field (Schramm, 1983, at p. 8; see also Schramm, 1997). Schramm himself held a bachelor's degree in political science, a Ph.D. in English literature, and had done postdoctoral research in a psychology department before he was appointed to a position in the area of journalism. His own career pattern thus exemplified the different roots of the field (Rogers, 1994).

The origins of communication studies in the United States date back to the beginning of the 20$^{th}$ century. Studies of World War I propaganda are often considered as the beginning of the field (Delia, 1987). The first study programs in *mass communication* in the U.S. emerged from this background after World War II. Doctoral degrees in *interpersonal communication* have been awarded by departments of speech since the 1930s (Rogers, 1994). However, despite gifted individuals who crossed existing boundaries and became leading figures in the field, the intellectual divide between mass communication—rooted primarily in the political sciences—and interpersonal communication with roots in (social) psychology has remained a serious drawback for theoretical advancement and institutional integration in communication studies.

A lack of communication (and consequently citation traffic) between the two sub-fields can be considered as a barrier to the identity formation of communication studies as a specialty in terms of scholarly journals and associations (Berger & Chaffee, 1988; Craig, 2003; O'Sullivan, 1999; Reardon & Rogers, 1988; Rogers & Chaffee, 1993). O'Keefe (1993) noted that the two subfields were increasingly becoming integrated in institutional terms. Whether the divide is still so dominant, however, has remained a point of discussion in this scientific community ever since (e.g., O'Sullivan, 1999; Putman, 2001). The *Journal of Communication*—a core journal of the field (see below)—devoted two special issues to this topic, entitled *The Ferment in the Field* (1983) and *The Future of the Field – Between Fragmentation and Cohesion* (1993), respectively. As recently as last year, a President's Message was issued in the *Newsletter of the International Communication Association* criticizing the current use of the ISI Subject Category "Communication" for the evaluation of communication studies (Rice & Putnam, 2007; cf. Bunz, 2005; Lauf, 2005).

Bibliometric research on communication studies as a specific field has sometimes criticized the ISI Subject Category "Communication", and worked on the basis of journal lists broader than the journals indexed in the *Social Science Citation Index* (Funkhouser, 1996, Rice *et al*., 1996). The ISI journal set contains a single Subject Category for "Communication." The scope note of this category specifies: "Communication covers resources on the study of the verbal and non-verbal exchange of ideas and information. Included here are communication theory, practice and policy, media studies (journalism, broadcasting, advertising, etc.), mass communication, public opinion, speech, business and technical writing as well as public relations." This is a very broad definition which goes beyond the two subdisciplines distinguished above. It includes journals such as *Written Communication* and *Telecommunications Policy* which do not—or no longer—fit within the parameters of communication studies as it has defined itself.

The scope note for social psychology ("covers resources on the behavior of the individual in a social context. Areas included are group processes, interpersonal processes, intercultural relations, personality, social roles, persuasion, compliance, conformity, sex roles, and sexual orientation.") indicates the overlap with communication studies more clearly than the one for political science ("covers resources concerned with political studies, military studies, the electoral and legislative processes, political theory, history of political science, comparative studies of political systems, and the interaction of politics and other areas of science and social science".) However, Pudovkin and Garfield (2002, at p. 1113) noted that journals are assigned categories by "subjective, heuristic methods." Although these categories may be sufficient for some purposes, the authors admit that "in many areas of research these 'classifications' are crude and do not permit the user to quickly learn which journals are most closely related."

Because of the well documented divide within the discipline, the specialty of communication studies provided us with an opportunity to test a new method for systematic delineation based on ideas generated in previous research efforts (Cozzens & Leydesdorff, 1993; Leydesdorff, 2004; Leydesdorff & Cozzens, 1993). Is it possible, in a straightforward way, to identify journal sets within a field as a subset of the grand matrix of aggregated journal-journal citation relations? Several authors have argued recently that this grand matrix is nearly decomposable (Leydesdorff & Rafols, 2009; Newman, 2006a and b; Rafols & Leydesdorff, in preparation; Rosvall & Bergstrom, 2008). Because of this property, one would expect that clear subsets can be extracted from relevant environments as specific densities of journals. More distant environments would not be expected to have much influence on local delineations. Is factor analysis able to distinguish the specific sets in an otherwise fuzzy environment?

The ISI Subject Categories are broadly defined for the purpose of bibliographic information retrieval, and they allow for overlap. However, we found in another context (Leydesdorff & Rafols, 2009; Rafols & Leydesdorff, in preparation) that the relations among the categories are statistically reliable despite potential errors in the individual attributions of journals to categories. Thus, the three Subject Categories (Communication, Social Psychology, and Political Science) can together be expected to constitute a wide net which contains more information than the set relevant for the delineation of communication studies as a specialty.

Within this larger set, one should be able to retrieve communication studies as one (or two?) subsets. The robustness of the subset(s) can further be tested by extending the environment. We shall do so in a later section by introducing the ISI Subject Category "Management" into the relevant environment first, and then by using the complete set of 1,865 journals of the *Social SCI*. Unlike these larger sets, however, the smaller sets allow us to visualize the development of the field by using animations.

**Methods and materials**

"Communication"—and not "communication studies"—is distinguished in the *Social Science Citation Index* as an ISI Subject Category. This category contained 45 journals in the *Journal Citations Report* (*JCR*) 2007. In addition to this category, we add the journals contained in the categories of Social Psychology (47 journals) and Political Science (93 journals). No journals in the resulting set are attributed to all three categories; the overlap between Political Science journals and Communication Studies involves three journals, and between the latter and Social Psychology only two journals. The overlap between Political Science and Social Psychology consists only of the journal entitled *Political Psychology*.

In other words, the three subsets are rather discrete in terms of the ISI Subject Categories, but in practice many scholars in communication studies maintain publication profiles—for example, in terms of co-authorship relations—showing their previous degrees in social psychology or political science, respectively. One could even question whether citation traffic across the interface within communication studies would be less dense than the citation traffic of the two subfields with their respective mother-disciplines. For example, in the list of cited references one would expect to find reflections of an author's educational background. However, these authors have on the one side left their mother-disciplines by choosing for communication studies. On the other side, their allegiance to the mother-disciplines may limit their visibility for the institutional colleagues on the other side of the divide as new audiences. By being embedded in a small sub-community on either side, one may harm one's chances of being cited.

We took all journals in the three categories as our set, and collected cross-citation data from the *JCR* at the aggregated journal level for each of the years 1994-2007. This provides us with (14) asymmetrical matrices of cited *versus* citing journals for each year. These matrices can be factor-analyzed along both axes (Q- and R-factor analysis). For the dynamic analysis, vector spaces were generated using the cosine as a similarity criterion (Ahlgren *et al.*, 2003; Salton & McGill, 1983). The cosine-normalized matrices were used as input to the animation using *visone* as described in Leydesdorff & Schank (2008).

The animations are available online at http://www.leydesdorff.net/commstudies/cited/index.htm for the cited dimension and http://www.leydesdorff.net/commstudies/citing/index.htm for the citing dimension, respectively.[1] They visualize the two main groups of journals in political science and social psychology, respectively, and the emerging set of journals considered as communication studies. The animations are based on minimizing the stress value both at each moment of time and over time. Stable linkages are not removed between years in order to optimize the conservation of a mental map upon inspection (Bender-deMoll & McFarland, 2006; De Nooy *et al.*, 2005; Moody *et al.*, 2005).

The integration in the group of communication studies journals is stronger in terms of the "being cited" patterns than in their citing patterns. As noted, we added in a next step the ISI Subject Category of "Management" containing 81 journals to the original set (of 179 journals), leading to a total set of 259 journals which can no longer meaningfully be visualized on a single screen. This larger set allows us to test the robustness of the analysis. Do we retrieve the same density for communication studies by using factor analysis? In a final step, the constructed set of journals in communication studies will be compared with a specific density in the grand matrix of all (1,865) journals included in the *Social Science Citation Index.* The constructed sets can be tested on their reliability as an indicator using Cronbach's α (Cronbach, 1951; Nunnaly & Bernstein, 1994).

[1] These animations use streaming. An alternative but slower format (for Windows Media Player) is provided at http://www.leydesdorff.net/commstudies/animations.htm.

## Results

Table 1 provides the numbers of the journals included in the initial analysis. The number of journals in political science is approximately twice as large as in communication studies or social psychology. While some journals are attributed to more than a single category, the number in the fourth column does not necessarily correspond to the sum of the first three columns.

| *Year* | *Political Science* | *Social Psychology* | *Communication* | *Three lists combined* |
|---|---|---|---|---|
| 1994 | 74 | 31 | 23 | 127 |
| 1995 | 68 | 34 | 28 | 128 |
| 1996 | 70 | 35 | 31 | 134 |
| 1997 | 73 | 40 | 36 | 144 |
| 1998 | 73 | 40 | 38 | 146 |
| 1999 | 76 | 41 | 43 | 155 |
| 2000 | 77 | 41 | 43 | 155 |
| 2001 | 78 | 43 | 43 | 158 |
| 2002 | 80 | 45 | 42 | 161 |
| 2003 | 78 | 46 | 44 | 162 |
| 2004 | 79 | 46 | 40 | 159 |
| 2005 | 84 | 46 | 42 | 166 |
| 2006 | 85 | 46 | 44 | 169 |
| 2007 | 93 | 47 | 45 | 179 |

**Table 1**: Number of journals in each of the three categories and the combined list.[2]

[2] While some journals are attributed to more than a single category, the number in the fourth column does not necessarily correspond to the sum of the first three columns.

As noted, the list for "Communication" as an ISI Subject Category (45 in 2007) can be considered as an overestimation of the specialty of "communication studies" since journals are included which belong to social psychology (e.g., the *Journal of Social and Personal Relations*) or other fields (e.g., discourse analysis; see below). This overestimation, of course, is legitimate in terms of the ISI's aim to assign categories for information disclosure across a wide variety of user groups.

Boyack *et al*. (2005, at p. 370) and Leydesdorff (2006, at pp. 611f.) found such overestimations in other fields as well. Boyack (personal communication, 14 September 2008) conjectured that the ISI Subject Categories are correct in approximately 50% of the cases. However, Leydesdorff & Rafols (2009) found that the remaining signal/noise ratio is sufficient for removing the noise by using factor analysis as a method for data reduction. Unlike the latter study based on the citation matrix among the categories, here we use the finer-grained matrix of cited and citing journals based on the sets listed in the last column of Table 1.

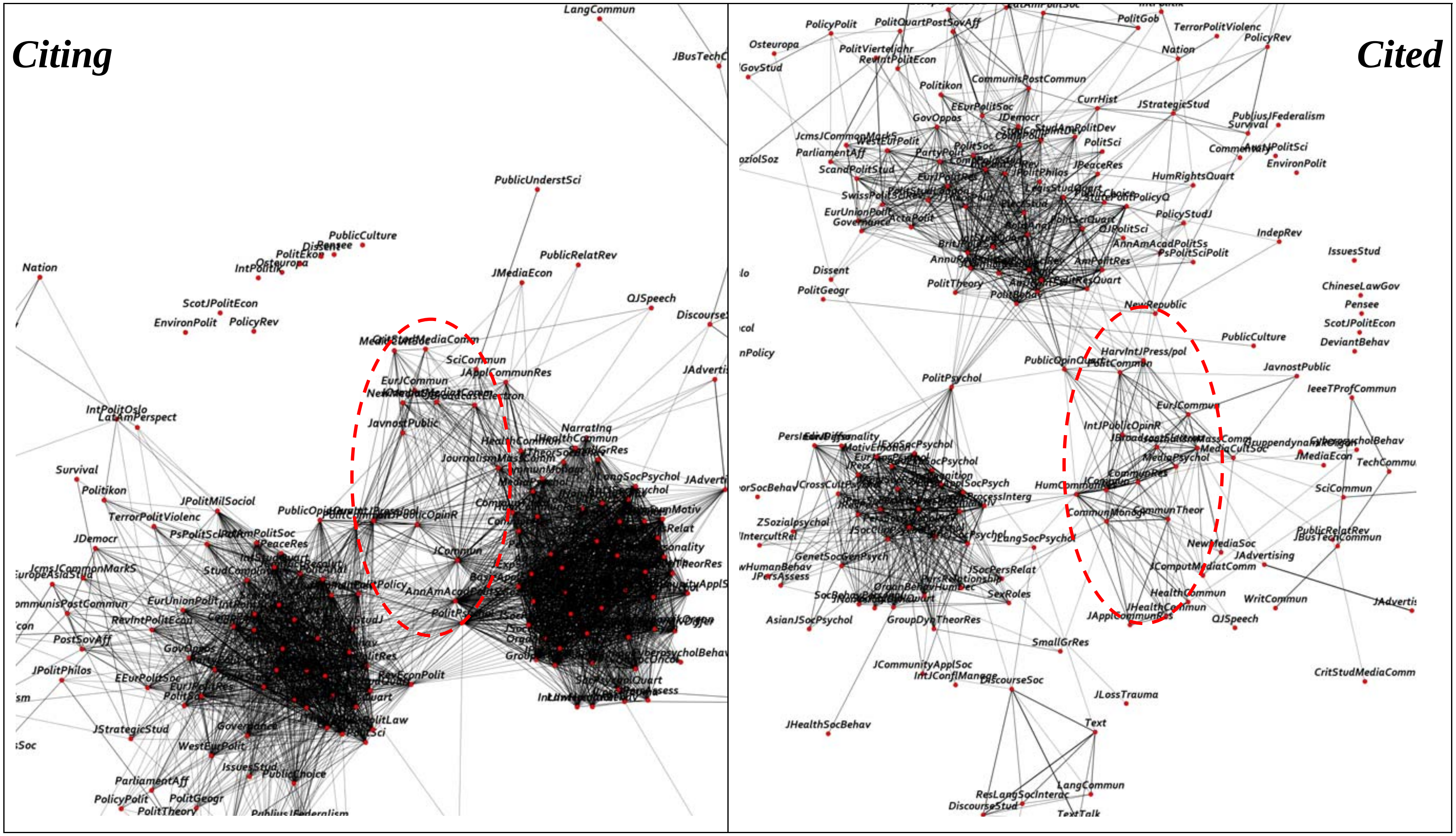

**Figure 1**: Citing (on the left) and Cited (on the right) patterns of 179 journals subsumed under the three ISI Subject Categories of "Communication", "Social Psychology," and "Political Science," in 2007. The circles indicate the "Communication" factor.

The animations show that an increased density of cross-citations among journals in communication studies emerged during the period under study (1994-2007), but the grouping is formed and visually distinguishable from the other two fields more clearly in the cited than in the citing dimension. Figure 1 shows the resulting configuration in 2007 for the "citing" and "cited" dimensions, respectively. We added auxiliary circles in order to indicate the specific domains of communication studies in these maps. The difference between the two pictures accords with our above conjecture that "citing" authors are inclined to remain closer to their disciplinary origins. The visibility of "communication studies" as a separate field in the "being cited" dimension is clearer, probably due to the visibility of institutionally collocated scholars in the relevant citation environments.

In other words, the set of journals in communication studies is visible as a single grouping more in its relevant contexts (of political science and social psychology journals), that is, in the cited dimension, than it is reproduced in terms of references provided by scholars publishing in communication studies itself. The specialty journals will therefore be delineated using the cited dimension of the journal-journal citation matrix. This difference between the cited and the citing dimension can be appreciated as a sign indicating that scholars working in communication studies perceive this field as interdisciplinary more than scholars in neighboring fields. These results accord with Rice *et al*. (1988) who using another methodology found similar differences among the cited and citing patterns of journals in communication studies, but not significantly different between the two major sub-groups (mass *vs*. interpersonal communications). Both subgroups in communication studies seem to import warrants for knowledge claims from other fields more than *vice versa*.

**More detailed analysis of the development (1994-2007)**

In order to make it possible to compare among different years and "cited" versus "citing," we chose to maintain a six-factor model across the years. In 2007, six factors explain 26.5% of the variance when using the "cited" patterns of the journals as variables, and 43.6% when using the (transposed) "citing" patterns of journals. The "citing" patterns are thus more structural than the "cited" ones. In other words, the divide is actively reproduced by this community of scholars.

Both "cited" and "citing," communication studies journals load highest on the third factor which, however, explains only 4.1% and 4.5% of the variance, respectively. In 1994, journals in communication studies loaded not on the third, but the fourth factor, which explained only 3.1% of the variance in this matrix. Thus, by using this method not only has the number of journals indicated as communication studies grown (from 14 in 1994 to 28 in 2007), but their internal coherence in terms of sharing a common variance in the relevant environments has also increased. The factor analysis enables us to distinguish this dimension in the data, and to list the journals with highest factor loadings (Table 2).

| | *1994 cited; N = 127* | *2006 cited; N = 169* | *2007 cited; N = 179* | *2007 citing; N = 179* |
|---|---|---|---|---|
| Cronbach's Alpha | 0.675 | 0.779 | 0.805 | 0.866 |
| 1 | *Commun Res* | *J Commun* | *J Commun* | *J Commun* |
| 2 | *J Commun* | *Commun Res* | *Commun Res* | *J Broadcast Electron* |
| 3 | *J Broadcast Electron* | *Hum Commun Res* | *Hum Commun Res* | *Journalism Mass Comm* |
| 4 | *Journalism Quart* | *J Broadcast Electron* | *J Broadcast Electron* | *Media Psychol* |
| 5 | *Hum Commun Res* | *Media Psychol* | *Journalism Mass Comm* | *Harv Int J Press/Pol* |
| 6 | *Public Opin Quart* | *Polit Commun* | *Media Psychol* | *Commun Res* |
| 7 | *Commun Monogr* | *Journalism Mass Comm* | *Commun Theor* | *Eur J Commun* |
| 8 | *Crit Stud Mass Comm* | *Commun Monogr* | *Polit Commun* | *Polit Commun* |
| 9 | *Q J Speech* | *Commun Theor* | *Commun Monogr* | *Hum Commun Res* |
| 10 | *Media Cult Soc* | *Eur J Commun* | *Int J Public Opin R* | *Commun Theor* |
| 11 | *Public Relat Rev* | *Int J Public Opin R* | *Eur J Commun* | *Javnost-Public* |
| 12 | *J Advertising* | *Harv Int J Press/Pol* | *Harv Int J Press/Pol* | *J Appl Commun Res* |
| 13 | *Small Gr Res* | *Health Commun* | *J Health Commun* | *New Media Soc* |
| 14 | *J Advertising Res* | *J Appl Commun Res* | *Health Commun* | *Health Commun* |
| 15 | *Telecommun Policy* | *J Health Commun* | *J Appl Commun Res* | *Int J Public Opin R* |
| 16 | *Commun Educ* | *New Media Soc* | ***J Comput-Mediat Comm*** | *Commun Monogr* |
| 17 | | *J Media Econ* | *Media Cult Soc* | *Media Cult Soc* |
| 18 | | *Media Cult Soc* | *New Media Soc* | *Sci Commun* |
| 19 | | *J Advertising* | *J Advertising* | *J Health Commun* |
| 20 | | *J Advertising Res* | *Sci Commun* | *Crit Stud Media Comm* |
| 21 | | ***Ann Am Acad Polit Ss*** | *Cyberpsychol Behav* | *J Comput-Mediat Comm* |
| 22 | | ***Public Culture*** | *Javnost-Public* | *J Media Econ* |
| 23 | | *Public Relat Rev* | *J Advertising Res* | *Cyberpsychol Behav* |
| 24 | | *Crit Stud Media Comm* | *Crit Stud Media Comm* | *Public Underst Sci* |
| 25 | | *Sci Commun* | *J Media Econ* | *J Advertising* |
| 26 | | *Javnost-Public* | ***Public Underst Sci*** | *Public Relat Rev* |
| 27 | | *Cyberpsychol Behav* | *Q J Speech* | *Q J Speech* |
| 28 | | *Q J Speech* | *Public Relat Rev* | |

**Table 2**: Listings of journals with highest factor loadings on the factor representing "Communication Studies" in different years. (Changes among the lists for 2006 and 2007 are boldfaced.)

Table 2 provides the listing of the journals loading on a factor representing communication studies in the years 1994, 2006, and 2007 using the "being cited" dimension. For 2007, the listing based on "citing" pattern is also added. By using boldface in Table 2 we indicated the two journals added in 2007 compared with 2006, and the two journals that were no longer classified as communication studies in 2007.

The *Journal of Advertising Research* is not part of the cluster in terms of its citing patterns. (We shall see below why.) The *Journal of Broadcasting and Electronics* has a much stronger profile in the citing dimension than in the cited one. The lists are quite similar, but the order differs between citing and cited. As noted, aggregated citation behavior and being cited patterns are different (cf. Rice *et al.*, 1988)

Cronbach's (1951) alpha can be used as a test for the reliability of the constructs in this environment. The list for 1994 *fails* the test using Nunnaly & Bernstein's (1994) criterion of $\alpha > 0.7$. However, the construct is reliable in 2007 in both the cited and citing dimension ($\alpha > 0.8$), and reliability was further improved between 2006 and 2007. The construct is more reliable "citing" than "cited" in accordance with the higher percentage common variance explained by the factor as noted above.

The network development may not have crystallized to the point where one can reach agreement among scholars in communication studies on whether this as a stable cluster. However, the lists are coherent and represent reliable constructs which are reproduced from year to year both citing and cited. The other factors are, as could be expected, journals in the area of social psychology (factor 1) and political science (factors 2, 4, and 5). The group of journals in the political sciences is further divided into specialty structures, such as American political science versus European or comparative studies of political systems.

Factor 6 can be distinguished both in 2006 and 2007 as a newly emerging cluster of journals that focus on texts and discourse analysis. This structure is visible in the right-hand picture ("cited") in Figure 1 at the lower end. In 2007, the two journals with the highest factor loadings on this newly emerging factor were *Discourse & Society* and *Discourse Studies.* The new dimension in the data was not yet recognizable using factor analysis in 2005. However, these journals are still attributed to the category of "Communication" in the ISI classification for 2007.[3]

**Further tests of the delineation**

One of the referees suggested that we could further test the stability of the cluster by adding the ISI Subject Category of "Management" to the previous list of three categories. Communication researchers often study organizations, and organizational communication may be cited in journals like *Organization Studies*, the *Journal of Management,* and the *Academy of Management Review* and *Journal*.[4]

The ISI Subject Category "Management" contains 81 journals in 2007, of which 80 were not included in the above used list of 179 journals. Using these 259 journals, 29 journals in communication studies have highest (positive) factor loadings on the sixth factor (which explains only 2.1% of the common variance). The new list covers the old list of 28 journals found above (Table 2), but the cited pattern of the journal *Public Opinion Quarterly* in this case additionally loads on the communication-studies factor slightly higher (0.341) than on the (third) political science factor (0.340). In the previous analyses, *Public Opinion Quarterly* was also interfactorially complex, but marginally sorted under the political science category on the basis of the criterion of using the highest factor loadings.

[3] *Discourse & Society* was also assigned to the categories "sociology" and "psychology, multidisciplinary."
[4] This referee also mentioned *Management Communication Quarterly*, but this journal has not been included in the *Social Science Citation Index.*

| | 2007 cited (*N* = 179) | Including 81 "Management" Journals (*N* = 259) | Factor 41 in the 50-factor solution using the *SoSCI* (*N* = 1858) |
|---|---|---|---|
| Cronbach's Alpha | 0.805 | 0.810 | 0.816 |
| 1 | *J Commun* | *J Commun* | *J Commun* |
| 2 | *Commun Res* | *Commun Res* | *Commun Res* |
| 3 | *Hum Commun Res* | *Hum Commun Res* | *Hum Commun Res* |
| 4 | *J Broadcast Electron* | *J Broadcast Electron* | *Journalism Mass Comm* |
| 5 | *Journalism Mass Comm* | *Journalism Mass Comm* | *J Broadcast Electron* |
| 6 | *Media Psychol* | *Media Psychol* | *Media Psychol* |
| 7 | *Commun Theor* | *Commun Theor* | *Commun Theor* |
| 8 | *Polit Commun* | *Polit Commun* | *Polit Commun* |
| 9 | *Commun Monogr* | *Commun Monogr* | *Commun Monogr* |
| 10 | *Int J Public Opin R* | *Int J Public Opin R* | *Int J Public Opin R* |
| 11 | *Eur J Commun* | *Eur J Commun* | *Eur J Commun* |
| 12 | *Harv Int J Press/Pol* | *Harv Int J Press/Pol* | *Harv Int J Press/Pol* |
| 13 | *J Health Commun* | *J Health Commun* | *J Health Commun* |
| 14 | *Health Commun* | *J Appl Commun Res* | *J Appl Commun Res* |
| 15 | *J Appl Commun Res* | *Health Commun* | *Health Commun* |
| 16 | *J Comput-Mediat Comm* | ***Public Opin Quart*** | *Public Opin Quart* |
| 17 | *Media Cult Soc* | *Media Cult Soc* | ***Am Behav Sci*** |
| 18 | *New Media Soc* | *J Comput-Mediat Comm* | *J Comput-Mediat Comm* |
| 19 | *J Advertising* | *New Media Soc* | *Media Cult Soc* |
| 20 | *Sci Commun* | *J Advertising* | *New Media Soc* |
| 21 | *Cyberpsychol Behav* | *Sci Commun* | *Javnost-Public* |
| 22 | *Javnost-Public* | *Cyberpsychol Behav* | *J Media Econ* |
| 23 | *J Advertising Res* | *Javnost-Public* | *Crit Stud Media Comm* |
| 24 | *Crit Stud Media Comm* | *J Advertising Res* | *Q J Speech* |
| 25 | *J Media Econ* | *Crit Stud Media Comm* | *Public Underst Sci* |
| 26 | *Public Underst Sci* | *J Media Econ* | *Public Relat Rev* |
| 27 | *Q J Speech* | *Q J Speech* | |
| 28 | *Public Relat Rev* | *Public Underst Sci* | |
| 29 | | *Public Relat Rev* | |

**Table 3**: Two further tests concerning the stability of the set of communication-studies journals.

Further extending the relevant environments, a grand journal-journal citation matrix can also be generated on the basis of all 1,865 journals included in the *Social Science Citation Index* 2007. Seven journals were cited fewer than five times in this database and were excluded from this analysis.[5] Using a 50-factor model (which explains 30.0% of the common variance),[6] for example, one can delineate a group of 26 communication journals loading on factor 41 (explaining 0.33% of the common variance). This list confirms the previous delineation (Table 3). However, the two journals about advertising research (the *Journal of Advertising* and the *Journal of Advertising Research*) are now sorted with a group of 35 marketing journals with their highest factor loadings on factor 21 (0.47%). *Science Communication* is in this larger context part of a group of 42 journals in library and information science which form factor 10 (0.79%). The journal *American Behavioral Scientist* shows interfactorial complexity in this model, with its major factor loading on the group of communication-studies journals. Otherwise, the lists are virtually similar. The addition of the management journals adds slightly to the reliability of the construct, since α = 0.810. The reliability is further enhanced by choosing the larger set of all 1858 journals as the relevant environment (α = 0.816).[7]

In summary, one can delineate a specific journal set in the cited dimension of an aggregated journal-journal citation matrix by using a few of the categories because this matrix is nearly decomposable into its constituent groups. This conclusion is a practical implication of the conclusion of previous studies about the ISI Subject Categories (Bensman & Leydesdorff, 2009; Leydesdorff & Rafols, 2009; Rafols & Leydesdorff, in preparation) that the ISI Subject Categories—because of the multiple assignments—tend to hide underlying structures in the data. On the basis of the near decomposability of the grand matrix, however, one can generate a shortlist representing a specific citation density by using a local environment. The list may remain uncertain at the margins, but provides a much more robust representation of a specific set than does the corresponding ISI Subject Category.

---

[5] These journals are: *Afrika Spectrum, Chinese Sociology and Anthropology, Journal of Women Politics & Policy, Nordic Psychology, Paedagogica Historica, Russian Politics and Law, Sociological Theory and Methods*.

[6] The choice of a 50-factor model (using SPSS version 15) is a bit arbitrary. The main reason for this choice was that given a set of 1858 journals, fifty groupings might lead to the right order of magnitude (20-50 journals) for making the comparisons under study.

[7] Removing the journal *Public Opinion Quarterly* improves reliability in both cases to α = 0.813 and 0.819, respectively.

Not surprisingly, the list based on the full matrix of 1858 journals included in the *Social SCI* has the highest reliability as indicated by Cronbach's alpha (α = 0.816). Given the fuzziness prevailing in the full, the factor represents specificity. Using this factor matrix (available at http://www.leydesdorff.net/commstudies/f50_cited.xls), one can, for example, distinguish a 10th factor which indicates 42 journals in the library and information sciences. This set is juxtaposed in Appendix 1 to the larger set of 56 journals classified under the ISI Subject Category "Information and Library Science." Thirty-eight of the cases overlap. By zooming in, one can further analyze the fine-structure of the set.

## Conclusions and discussion

In the case of new and interdisciplinary developments, relevant ISI Subject Categories may be available as in the above case, but precise delineation has remained a problem. The Subject Categories are attributed by the ISI on bibliographic grounds and not for the purpose of journal or research evaluation. In the case of communication studies, for example, the delineation has remained heavily debated within the relevant community. However, in recent years a core set seems to be stabilizing. Journals like the *International Journal for Public Opinion Research* and *Political Communication* now belong to this core set, but *Public Opinion Quarterly* is identified as a journal at the interface with political science.[8]

The distinction from social psychology has become sharper than that from the political sciences (which themselves are also more complex in terms of using a variety of paradigms). Although still located between social psychology and communication studies in 2007, *Human Communications Research* has since 1994 been a core journal of communications studies. In the listing of Table 2, its position has improved from fifth (in 1994) to third (in 2007) in terms of factor loadings. In the animations, however, the demarcation of communication studies from social psychology is only clear in the "cited" patterns (at http://www.leydesdorff.net/commstudies/cited/index.htm).

[8] In terms of the ISI Subject Categories, *Public Opinion Quarterly* is multiply assigned to the categories "Communication," "Political Science," and "Social Science, Interdisciplinary."

The prevailing impression remains that "communication studies" cannot yet be indicated as a stable set, but the set is in transition towards the establishment of a specialty of its own. Yet, not being completely internally clustered doesn't mean that a set is unstable. The differentiation with this field seems to be structural since reproduced by the publishing scholars from year to year. Many communication departments intentionally hire and teach, and conduct research in both mass media and interpersonal communications (as well as in other specializations, such as organizational, policy, new media, etc.) specifically to cover the subareas of the field.

Journals need to have an identity, both for publishing and citing authors, but certainly for readers. So as disciplines develop, and even stabilize and strengthen their identity, they tend to have more journals, each more specialized. Complete and undistinguished cross-citation patterns could in this case mean not a comprehensive and stable theoretical identity, but rather might imply a lack of distinctions or meaningful concepts. The internal divide between interpersonal and mass communication seems to be functional to the intellectual reproduction of this field as a social unity, with journal articles citing from social psychology or political studies, while from the outside communication studies can increasingly be perceived as an independent source of knowledge claims (Leydesdorff & Van den Besselaar, 1997; Whitley, 1984).

The aggregated "citing" patterns are both more divided than the "cited" ones and these patterns are more structural—that is, a higher percentage of the variance is explained by the corresponding factors. At a sociological level, one can wonder whether a field that is still so divided (after more than two decades) will in the near future be able to continue to absorb the large number of students and scholars who are striving for a career (and tenure) in it. Obviously, the journal system is conservative and has built-in resistances to change, among them the admission procedures of the ISI (Garfield, 1990; Testa, 1997). However, we expect that the cluster of journals in communication studies will gain in terms of further coherence as more scholars contribute to these specialist journals with degrees in communication studies itself instead of backgrounds in the political sciences or social psychology.

Note that our lists of journals for "communication studies" (in Table 3) are more restricted than the ISI Subject Category of "Communication." It seems legitimate to us that the ISI should cast a wider net for the purpose of information retrieval. For the evaluation of research and journals the smaller set may be more appropriate because of its stronger focus and its legitimacy in terms of journal-journal citation analysis. In national contexts, one may wish to extend this list with relevant journals in the respective languages (Lepori & Probst, forthcoming). Although the lists remain fluid from year to year, the method submitted is rather straightforward in each year and allows for the initial journal delineation in cases where one would expect a set to be considerably fuzzy.

**Acknowledgement**

We are grateful to extensive referee comments on previous versions of this paper.

return

**Appendix I**. Comparison for Library & Information Science.

| 42 journals with highest factor loadings on factor 10 of the aggregated factor matrix in 2007.[9] | 56 journals attributed to the ISI Subject Category "Information & Library Science" in the JCR 2007. |
|---|---|
| *Annu Rev Inform Sci* | *Annu Rev Inform Sci* |
| *Aslib Proc* | *Aslib Proc* |
| *Coll Res Libr* | *Coll Res Libr* |
| *Comput Linguist* | |
| *Econtent* | *Econtent* |
| *Electron Libr* | *Electron Libr* |
| | *Gov Inform Q* |
| *Health Info Libr J* | *Health Info Libr J* |
| | *Inform Manage-Amster* |
| *Inform Process Manag* | *Inform Process Manag* |
| *Inform Res* | *Inform Res* |
| *Inform Soc* | *Inform Soc* |
| | *Inform Syst J* |
| | *Inform Syst Res* |
| *Inform Technol Libr* | *Inform Technol Libr* |
| | *Int J Geogr Inf Sci* |
| | *Int J Inform Manage* |
| *Interdiscipl Sci Rev* | |
| *Interlend Doc Supply* | *Interlend Doc Supply* |
| *J Acad Libr* | *J Acad Libr* |
| | *J Am Med Inform Assn* |
| *J Am Soc Inf Sci Tec* | *J Am Soc Inf Sci Tec* |
| | *J Comput-Mediat Comm* |
| *J Doc* | *J Doc* |
| | *J Glob Inf Manag* |
| | *J Health Commun* |
| *J Inf Sci* | *J Inf Sci* |
| | *J Inf Technol* |
| *J Libr Inf Sci* | *J Libr Inf Sci* |
| | *J Manage Inform Syst* |
| *J Med Libr Assoc* | *J Med Libr Assoc* |
| *J Scholarly Publ* | *J Scholarly Publ* |
| *J Urban Technol* | |
| *Knowl Organ* | *Knowl Organ* |
| | *Law Libr J* |
| *Learn Publ* | *Learn Publ* |
| *Libr Collect Acquis* | *Libr Collect Acquis* |
| *Libr Inform Sc* | *Libr Inform Sc* |
| *Libr Inform Sci Res* | *Libr Inform Sci Res* |
| *Libr J* | *Libr J* |
| *Libr Quart* | *Libr Quart* |

[9] The factor matrix is available at http://www.leydesdorff.net/commstudies/f50_cited.xls .

| | |
|---|---|
| *Libr Resour Tech Ser* | *Libr Resour Tech Ser* |
| *Libr Trends* | *Libr Trends* |
| *Libri* | *Libri* |
| | *Mis Quart* |
| *Online* | *Online* |
| *Online Inform Rev* | *Online Inform Rev* |
| *Portal-Libr Acad* | *Portal-Libr Acad* |
| *Program-Electron Lib* | *Program-Electron Lib* |
| *Ref User Serv Q* | *Ref User Serv Q* |
| *Res Evaluat* | *Res Evaluat* |
| | *Restaurator* |
| *Sci Commun* | |
| *Scientist* | *Scientist* |
| *Scientometrics* | *Scientometrics* |
| *Serials Rev* | *Serials Rev* |
| *Soc Sci Comput Rev* | *Soc Sci Comput Rev* |
| | *Soc Sci Inform* |
| | *Telecommun Policy* |
| | *Z Bibl Bibl* |

return